\documentclass[twocolumn,letterpaper]{IEEEAerospaceCLS}  

\usepackage[]{latexsym,amssymb,amsfonts,amsmath,amstext,times} 

\usepackage[]{graphicx}    
\usepackage{float}
\newcommand{\ignore}[1]{}  

\begin{document}
\title{GEOSCAN: Global Earth Observation using Swarm of Coordinated Autonomous Nanosats}

\author{
Changrak Choi\\
Jet Propulsion Laboratory\\
California Institute of Technology\\
Pasadena, CA 91109\\
changrak.choi@jpl.nasa.gov
\and 
Anthony B. Davis\\
Jet Propulsion Laboratory\\
California Institute of Technology\\
Pasadena, CA 91109\\
anthony.b.davis@jpl.nasa.gov
%
\thanks{\footnotesize \copyright 2021 California Institute of Technology. All rights reserved.}
}

\maketitle

\thispagestyle{plain}
\pagestyle{plain}

\maketitle

\thispagestyle{plain}
\pagestyle{plain}

\begin{abstract}
The climate crisis we are facing calls for significant improvements in our understanding of natural phenomena, with clouds being identified as a dominant source of uncertainty. To this end, the emerging field of 3D computed cloud tomography (CCT) aims to more precisely characterize clouds by utilizing multi-dimensional imaging to reconstruct their outer and inner structure. In this paper, we propose a future Earth observation mission concept, driven by the needs of CCT, that operates constellation of NanoSats to provide multi-angular, spectrally-resolved, spatial and temporal scientific measurements of natural atmospheric phenomena. Our proposed mission, GEOSCAN, will on-board active steering capability to rapidly reconfigure networked swarm of autonomous Nanosats to track evolving phenomena of interest, on-demand, in real-time. We present the structure of the GEOSCAN constellation and discuss details of the mission concept from both science and engineering perspectives. On the science side, we outline the types of remote Earth observation measurements that GEOSCAN enables beyond the state-of-the-art, and how such measurements translate to improvements in CCT that can lead to reduction in uncertainty of the global climate models (GCMs). From the engineering side, we investigate feasibility of the concept starting from hardware components of the NanoSat that form the basis of the constellation. In particular, we focus on the active steering capability of the GEOSCAN with algorithmic approaches that enable coordination from new software. We identify technology gaps that need to be bridged and discuss other aspects of the mission that require in-depth analysis to further mature the concept.
\end{abstract}

\tableofcontents

\section{Introduction}

The recently released IPCC report 2021 exposes failure of mankind to act on the climate crisis that affects our everyday lives \cite{ipcc2021climate}. Our home, Planet Earth, is getting warmer, sea level is rising, dry and wet regions will likely become drier and wetter, and extreme weather phenomena are expected to intensify. This situation puts a huge societal urgency on Earth scientists to improve our understanding of atmospheric, oceanic, and land-surface processes as well as their complex interactions and feedbacks, with clouds being identified as a dominant source of uncertainty \cite{ipcc2021climate}. The path forward calls for innovative Earth observation solutions capable of acquiring multi-angular, spectrally-resolved, spatial and temporal observations to reduce significantly cloud-related uncertainties in global climate models (GCMs). The Earth observation solution can also be utilized to obtain accurate 3D spatio-temporal maps of cloud-like natural atmospheric phenomena such as dust storms, wildfires, volcanic ash emissions, all of which are hazards to be closely monitored.


{\it State-of-the-art}

Currently, NASA's cloud remote sensing capability is aboard its Terra and Aqua flagship satellite missions from the EOS era (late 1990s, early 2000s) and on CloudSat, like Aqua, in the A-train.
Also in the A-train constellation are PARASOL and CALIPSO \cite{stephens2018cloudsat}, respectively, an ESA mission and a joint NASA-ESA one.
Terra carries the Multi-angle Imaging Spectro-Radiometer (MISR) and MODerate resolution Imaging Spectrometer (MODIS), with the latter also on Aqua.
MISR is unique in its multi-angle take on the vertical dimension of clouds using passive imaging over a $\sim$400~km swath with 275~m pixels in the red channel; its other Visbile-Near-IR (VNIR) channels add little for clouds.
By contrast, MODIS has a swath $\sim$2330~km and boasts several spectral channels that are informative about cloud properties, ranging across the VNIR, Short-Wave-IR (SWIR), and Thermal-IR (TIR) spectrum at resolutions ranging from 250~m to 1~km.
CloudSat carries the Cloud Profiling Radar (CPR), a mm-wave radar sensitive to cloud droplets and drizzle, while CALIPSO carries the Cloud-Aerosol Lidar with Orthogonal Polarization (CALIOP), a polarized backscatter lidar.
These active sensors sample the vertical structure the clouds (top layers only for CALIOP), but only along the sub-satellite track.
PARASOL carries the 3$^\text{rd}$ incarnation of the Polarization and Directionality of the Earth's Reflectances (POLDER-3) sensor, a polarization-capable multi-angle imager with $\sim$10~km pixels.
To these assets in Low-Earth Orbit (LEO), we add the Earth Polychromatic Imaging Camera (EPIC) on the NASA/NOAA DSCOVR platform located near the Lagrange-1 point, $\sim$1.5\,10$^6$~km from Earth roughly in the direction of the Sun.
EPIC images almost the entire sunlit hemisphere at UV-VNIR wavelengths with $\sim$10~km pixels.

All of the above-mentioned sensors have a way of determining cloud top height, either via stereoscopy (MISR), using echo-location (CPR and CALIOP), capitalizing on thermal stratification (MODIS), or exploiting O$_2$ absorption (POLDER-3 and EPIC).
CPR and CALIOP data have been successfully merged to provide full-column cloud characterization, largely by filling in for each others blind spots.
POLDER-3 has broken new ground in the determination of cloud-top microphysical properties using polarization observations in the rainbow region of the scattering angle (not always available, unfortunately),
That said, the two MODIS sensors are the real workhorses in terms of generating cloud products from the Level~1 radiances they measure.
Every image (known as a ``granule'') is processed at the 1~km scale into cloud optical thickness and cloud particle effective radius using a straightforward bi-spectral technique \cite{nakajima1990determination}.
However, deep down, the forward model used in the algorithm assumes that the cloud's geometry is that of an infinite plane-parallel slab, thus enabling the computation of radiances using 1D radiative transfer (RT) code.
Understandably, this operational algorithm works reasonably well on single-layer stratiform clouds, at least at some distance from their boundaries.
At the opposite end of possible cloud types, vertically-developed cumuliform clouds driven by shallow or deep convective dynamics are underserved by the current algorithms at Level~2.
CPR can probe big-enough convective clouds but, again, their horizontal sampling is poor by comparison with passive imagers.

This shortcoming in the remote sensing of clouds caught the attention of the National Academies when collating their 2017 Decadal Survey (DS17) \cite{NASEM2018}, starting from how much it is holding back atmospheric science, especially in climate prediction using GCMs.
Indeed, convection along with the 3D clouds and intense precipitation that it can generate are all sub-gridcell processes that are currently not well understood.
In fact, far better comprehension of cloud-scale processes is required to model accurately their impact at the $\sim$100~km scale of a GCM gridcell, and thus capture cloud feedback mechanisms in the climate system.
Moreover, aerosols interact with clouds throughout their lifecycle in ways that are also very poorly understood, thus compounding the uncertainty in GCM predictions.
In order to orchestrate coordinated scientific action to bridge this knowledge gap, DS17 defined the Aerosol-Cloud-Convection-Precipitation or ``ACCP'' class of Designated Observables, meaning high-value/high-priority and, in turn, ACCP spawned the new Atmospheric Observing System (AOS) cluster of satellite missions for the reminder of this decade.

Due to the need to ensure continuity of the ``Program-of-Record'' established with current satellite missions, technological and algorithmic advances in ACCP/AOS will be largely incremental.
However, the challenges posed by the complex geometry and internal variability of vertically-developed convective clouds call for a radical departure from current bi-spectral 1D-RT-based retrievals.
That is precisely the motivation for recent developments in computed cloud tomography (CCT), which is predicated on 3D RT and powerful new methods for addressing large ill-posed inverse problems. 
CCT has been successfully demonstrated using synthetic multi-angle/multi-spectral/multi-polarization images (with known truth) based on a process-level dynamical modeling of clouds using a Large-Eddy Simulation (LES) code \cite{Levis_etal2015,Levis_etal2017,Levis_etal2020,martin2014adjoint,martin2018demonstration,doicu2021cloud}. 

CCT was also convincingly applied to real cloud observations \cite{Levis_etal2015,Levis_etal2021} by the Airborne Multi-angle Spectro-Polarimetric Imager (AirMSPI) \cite{diner2013airborne}.
Both synthetic and real-world data were sampled at 20- to 40~m resolution, which is well-suited for the embedded SHDOM 3D RT model \cite{evans1998spherical}.
In turn, that breakthrough in cloud remote sensing---now viewed as a 3D imaging problem---motivated the CloudCT mission \cite{schilling2019cloudct}.
CloudCT is a formation of $\sim$10 cubesats in a ``string-of-pearls'' configuration \cite{tzabari2021cloudct} designed to be a technology demonstration for doing from space what was done using LES and AirMSPI, including the use of its polarization channels \cite{Levis_etal2021}.

{\it Our Approach}

\begin{figure*}[t!]
	\centering
	\includegraphics[width=1.0\linewidth]{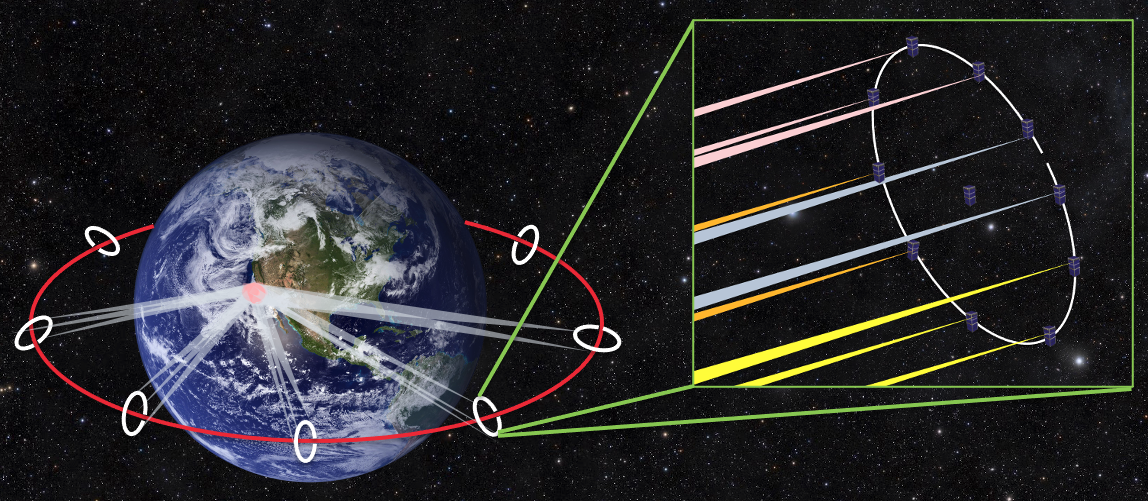}
	\caption{Our proposed mission will involve a sustainable networked swarm of autonomous NanoSats that can actively reconfigure to provide superior high-resolution 3D scientific measurements of Earth phenomena, on-demand, in real-time. The constellation will consist of tens of fleets (white) that are following one another in multiple string-of-pearls formation in low Earth orbit, separated by distances of hundreds of kilometers. Each fleet (pearl) will then consist of tens of NanoSats that are formation flying in close proximity (hundreds of meters) and carry heterogeneous scientific instruments. The string-of-pearls formation of fleets will enable multi-angular and temporal observations of a phenomena of interest. The close-proximity formation of NanoSats within fleet will provide spectral diversity needed for CCT reconstruction, with each type of scientific instruments (shown in different color) carried by multiple NanoSats.}
	\label{fig:mission_concept}
\end{figure*}


In this paper, we propose a future Earth observation mission concept, driven by the needs of the computed cloud tomography (CCT), that utilizes a constellation of NanoSats to provide such solution. It thus follows in the steps of CloudCT, but with an order-of-magnitude more agents and off-track pointing agility to better observe convective 3D clouds and other 3D cloud-like atmospheric phenomena (e.g., wildfire smoke and volcanic ash plumes).

Distributed spacecraft missions (DSM) involving constellations of spaceborne small satellites is gaining renewed interest as it effectively provides much needed multi-perspective observation from multiple vantage points. The recent advances in hardware miniaturization, ranging from instruments, on-board processors, navigation sensors, propulsion and communication units, have matured to a point where it is possible to deploy a large-scale constellation that consist of swarms of extremely small space-borne sensing nodes (e.g., CubeSats) in Earth orbit.

Our proposed mission will involve a sustainable networked swarm of autonomous NanoSats that can actively reconfigure to provide superior high-resolution 3D scientific measurements of Earth phenomena, on-demand, in real-time. The constellation will consist of tens of fleets that are following one another in multiple string-of-pearls formation in low Earth orbit, separated by distances of hundreds of kilometers. Each fleet will then consist of tens of NanoSats that are formation flying in close proximity (hundreds of meters) and carry heterogenous scientific instruments. The string-of-pearls formation of fleets will enable multi-angular and temporal observations of a phenomena of interest. The close-proximity formation of NanoSats with fleet will provide spectral diversity needed for CCT reconstruction.

The paper discusses details of the proposed mission concept from both science and engineering perspectives. Science side, we outline what types of remote Earth observation measurements are possible with the proposed mission concept (which are currently not possible) and how such measurement translates to improvements for CCT and, from there, reduction in uncertainty of GCMs. Engineering side, we investigate feasibility of the concept from both hardware and software components and identify possible gaps to be addressed. Recent advances in hardware miniaturization have matured to a point where it is possible to deploy a large-scale constellation that consists of swarms of extremely small space-borne sensing nodes (e.g., NanoSats) in Earth orbit. However, there are gaps on the software side, especially relating to the coordination and operation of complex constellation, that will be closely examined.

Our proposed distributed spacecraft mission concept has the potential to enable rapidly adaptive and agile constellation that can dynamically reconfigure its sensing nodes to observe, on-demand, natural phenomena of highest interest. This will enable real-time tracking of evolving phenomena, such as convectively-driven cloud complexes (hence heavy rain, floods, and snow storms) and dense plumes of wildfire smoke or volcanic ash through rapid and dynamical response that is not possible with the state-of-the-art methodologies.

{\it Outline}

The paper is organized as follows: An overview of the proposed mission concept and its details are given, with mission parameters identified in Section 2. Science aspects of the mission concept, ranging from on-board instrumentation to observation measurements, are discussed in relation to CCT improvements and GCM uncertainty reduction in Section~3. The following Section 4 looks at the engineering side, with feasibility study of the proposed mission and technology gaps/needs identified. Section 5 discusses various aspects of the mission concept that are important to consider and will require in-depth analysis to further concretize the concept. The paper concludes with a summary in Section 6.

\section{Mission Concept}

Our proposed mission, Global Earth Observation using Swarm of Coordinated Autonomous NanoSats (GEOSCAN), strives to be an active networked system that is far different from the currently existing and planned large satellite constellations that are passive in nature. The state-of-the-art satellite constellations (e.g., Star-Link) strive toward global coverage with satellites that are passively in orbit to cover certain pre-defined regions on Earth. GEOSCAN, on the other hand, does not aim for a global coverage. Instead, it strive towards having active agility that can precisely point multiple distributed high-res narrow field-of-view/swath scientific sensors to a particular location of interest on the fly, in real-time.  This requires very close coordination among the satellites with active pointing adjustment capability in a networked setting — and this coordination is done autonomously without ground-in-the-loop guidance leveraging upon recent advances in the Artificial Intelligence and Machine Learning.

\subsection{Constellation Structure}

The schematic of the GEOSCAN concept is depicted in Fig.\ref{fig:mission_concept}. GEOSCAN proposes active autonomous and networked NanoSats in the order of a hundred, which far exceeds current similar state-of-the-art that are in the order of few to ten at most \cite{bandyopadhyay2016review}. The closest to swarm network proposed in GEOSCAN is CloudCT that involves around 10 NanoSats, in a same orbital track (i.e., ``string-of-pearls'' formation flying). However, with only 10 active satellites, the CloudCT mission is limited to providing multi-directional measurements only under the chief orbit for 3D reconstruction of a cloud field. With the scale of a hundred of NanoSats, the GEOSCAN network system attains a critical mass that enables truly focused multi-angular and temporal as well as multi-spectral view of natural phenomena. Moreover, it allows for sustainable distributed sensor coverage that is robust to failures and responsive to time-sensitive dynamical natural phenomena.

The GEOSCAN constellation will be situated in low-earth orbit (LEO), flying closer to the Earth than the current ISS orbit. As shown in Fig.~\ref{fig:mission_concept}, GEOSCAN will consist of tens of fleets that are following one another in multiple string-of-pearls formation. The figure depicts one of such string-of-pearls (red) with fleets of spacecraft (white) riding along the same orbital track on a different phase to one another. Each fleet (white) will provide different angular measurement of a natural phenomena from one another at a given time. The fleets, as a whole, will provide multi-temporal measurement of a natural phenomena as it travels along the orbit in time. The proposed GEOSCAN has an order of 10 fleets placed in one string-of-pearls orbital track that are hundreds of kilometers apart. The optimized parameter value for how many fleets to be placed in one string-of-pearls orbital track and their respective phase difference needs tuning based on the science requirements. The total number of orbital tracks with string-of-pearls in the GEOSCAN constellation can also be varied and the choice of orbits will need to be optimized to maximize science observations.

Each fleet will then consist of tens of NanoSats that are formation flying in a wheel configuration (white) relative to the orbital track of the string-of-pearls (red). A fleet will consist of one NanoSat that acts as a chief and the rest of the NanoSats acting as deputies that orbits in a wheel formation around the chief. The wheel radius can be varied from one hundred meters to one kilometers, and the NanoSats within the wheel will be following one another in close-proximity on an identical relative orbit with respect to the chief. The fleet, as a whole, will provide spectral diversity of heterogeneous scientific measurements needed for CCT reconstruction. Each NanoSat, however, will be carrying one type of scientific instrument only and the cardinality of NanoSat in the fleet will depend on the different types of scientific measurement needed as well as spectral diversity required to reduce the uncertainty in the science model below the desired threshold. Note that the chief within a fleet will be primarily responsible for orbital tracking and communication, and does not take scientific measurement itself.




\subsection{Adaptive Pointing and Coordinated Observation}
The GEOSCAN constellation will heavily rely on autonomy for its operation as traditional ground-in-the-loop guidance cannot directly be scaled to manually commanding each NanoSats in the swarm. One of the biggest enabling feature will be capability to autonomously coordinate sensor pointing among available NanoSats within a fleet that form wheel, as well as across different fleets on the string-of-pearls formation. When a natural phenomena to be focused is given, GEOSCAN will generate optimal attitude plan that coordinates pointing of distributed sensors on-board different NanoSats to maximize scientific value of the measurements. This will involve constant selection/de-selection of team members within the swarm and subsequent attitude adjustments in real-time to maintain desirable multi-angle views on a particular natural phenomenon at focus. The optimal swarm attitude plan will also satisfy the required precision and agility while working with limited energy and actuation constraints. 

In routine operation, GEOSCAN will primarily deliver high-resolution 3D mapping of clouds. As previously discussed, clouds play critical role in Earth climate by controlling the energy fluxes and regulating the water cycle. The ability to accurately quantify the 3D macro and micro-physical characteristics of warm convection within clouds provide key insight that can greatly reduce the major uncertainties in the climate science \cite{ipcc2021climate} as well as challenges in numerical weather prediction. The impact of such improvement in weather forecasting can never be emphasized enough as it tremendously benefits every nation in the world to every person in their daily lives.

In special operation, GEOSCAN will switch focus to a particular natural phenomenon of the highest interest/concern on-demand at a given time. Examples of such natural phenomena will include rapidly spreading wild fire, dynamically evolving hurricanes, transitioning volcano plumes and many more natural hazards that require close monitoring. The real-time 3D spatio-temporal map of these natural hazards will immensely help to generate accurate plans to mitigate negative impacts on the affected regions, and potentially saving thousands of lives around the world.


\section{Science and Applications}

\subsection{Mission Objective}

In short, GEOSCAN has one primary science goal and one application area, both with major societal impacts.

The primary science goal is to enable better forecasting of Earth's future climate under a wide range of scenarios for greenhouse gas emissions and pollution controls.
This progress will be through more accurate parameterizations of clouds in convective dynamical regimes, a long-identified source of uncertainty in GCMs.
The enabling Level-2 product is CCT output: a gridded description of the inner and outer structures of vertically-developed clouds that can be directly compared with LES model output for LES validation.
It is widely believed that LES modeling of cloud-scale processes will be the computational fluid dynamics (CFD) framework from which the improved sub-grid GCM parameterizations will emerge.

The application area is the 3D reconstruction using CCT of cloud-like atmospheric phenomena involving other materials than condensed water and that are closely associated with natural hazards.
We can specifically call out opaque plumes of wildfire smoke and volcanic ash near their sources.
CCT techniques applied to such non-water ``clouds'' will provide an unprecedented quantification of these aerosol plumes at their origins, when they are still optically thick.
In the case of wildfires, the primary stakeholders are first responders who can better coordinate knowing more about the smoke production rate.
In the case of volcanoes, they are the airspace control professionals that can communicate hazards to airline pilots in a more timely and quantitative way.
In both cases, the GEOSCAN CCT products can provide key constraints (better initial conditions) on the prediction of long-range transport of the hazardous aerosols.

\subsection{Computed Cloud Tomography from Space}

CCT from space can follow two paths defined by the fore-optics used to focus incoming light onto a notional focal-plane array with 1~Mpix capacity: either 1) use high-enough spatial resolution (a few 10s of meters) to provide Level-1 multi-view images that can be fed directly into existing and validated CCT code, or minor variations thereof, and pay the price in swath width (at most, a few 10s of km); or else 2) use low-enough spatial resolution (a few 100s of meters) for the Level-1 radiance fields so that the swath width is several 100s of km, enough to capture complex meso-scale convective cloud systems, to better serve the science goals, and pay the price in algorithm development, knowing that first steps toward CCT with bigger pixels and more massive clouds have already been taken, at least in simulation \cite{forster2021toward}.

The former path is, understandably, the one chosen by the CloudCT mission in view of the short timeline for sensor and algorithm development, prior to launch planned for 2022.
With its longer timeline and broader choice of sensors, GEOSCAN can in fact take both paths.
The sensors/NanoSats with the smaller swath basically provide a 10$\times$ zoom into the details not captured by their wide-swath counterparts.

\subsection{Science Payload}

To be clear, in GEOSCAN, imagers with viewing angle diversity are absolutely essential to the Level-1 data processing using CCT.
That is realized in CloudCT as well as in GEOSCAN by NanoSats flying in a ``string-of-pearls'' formation along the same LEO orbit (say, $\sim$400~km altitude).
The orbital inclination should be at least $\sim$60$^\circ$ in order to cover all of the tropics and mid-latitudes where convective dynamical regimes occur.
If each ``pearl'' is separated by 100 to 300~km, increasing with distance from the central pearl, the formation will achieve the desirable range of view angles using $\sim$10 pearls.
The main difference between CloudCT and GEOSCAN is that, in the former, each pearl is a single NanoSat while in GEOSCAN it is a sub-formation of NanoSat flying in a tight ``cartwheel'' formation.
Each NanoSat (not tasked specifically with computation or communication) in any given wheel will carry a single sensor and, together, they ensure the required spectral diversity to the observation system as a whole.
Each sensor should be as simple as possible to keep the initial cost as low as possible, as well as the cost of replenishing the formation when a sensor/NanoSat fails, thus prolonging the duration of the GEOSCAN mission.

We can separate the sensors into ``desirable'' (a.k.a. baseline) and ``required'' (a.k.a. threshold) categories.
Starting with the later that constitute the minimum set for the mission to succeed, we have
\begin{itemize}
\item
CCD-type camera, standard RGB filters, high-resolution / narrow-swath fore-optics, vicarious calibration; this sensor will distinguish easily between clouds, smoke, ash and other types of aerosol; can downlink only red channel data if only clouds over water are being observed.
\item
Same as above, but with moderate spatial resolution and wide swath.
\item
SWIR camera (MODIS channels at 1240, 1640 and 2130~nm) with fore-optics set for the narrow swath; SWIR focal plane arrays typically have $\sim$1/2 as many pixels as their VNIR counterparts, and accordingly larger pixels; this sensor is required to gain sensitivity to cloud particle size at different depths into the cloud; it can also help in aerosol type discrimination.
\item
Same as above but for the wide swath captured at moderate spatial resolution.
\end{itemize}
Additionally, we require extra nadir-looking cameras in first off-nadir wheels on both sides of nadir wheel, and two extra off-nadir cameras in the nadir wheel; red channel only; high-resolution/narrow-swath only; this translates to three nadir looks at any given cloud with somewhat less than a minute time delay between each one, but also simultaneous with an off-nadir look.
From there, four stereo pairs are formed and by tracking cloud features between them, hence a direct measurement of cloud top vertical motion, a key quantity in convective dynamics.

As a first tradeoff, the required SWIR cameras may be placed only in the nadir- and most oblique-looking cartwheels, thus preserving the critical insights into the height dependence of the cloud microphysics.
Desirable sensors, also in just the nadir- and most oblique-looking cartwheels, are:
\begin{itemize}
\item
Extra RGB cameras, like the required ones, but with $\pm$45$^\circ$ polarization filters; these sensors will \emph{occasionally} capture light in the rainbow-to-glory range of scattering angles, where polarization state is highly sensitive to the cloud droplet size distribution (mean \emph{and} variance); this opens up a vista into complex microphysical processes at the interface of cloudy air and clear air (containing aerosols that may be affecting the cloud).
\item
TIR cameras in the most transparent atmospheric window, with vicarious calibration; narrow-swath only, typically with lesser spatial resolution than in the SWIR; these sensors will reveal the vertical thermal structure of the clouds and aerosol plumes, which is highly desirable to understand the prevailing convective dynamics.
\end{itemize}
Lastly, it is desirable to have cameras in all the cartwheels with an O$_2$ (A-band, 765~nm) absorption channel, a continuum channel, and an H$_2$O channel (936~nm) to perform multi-angle differential optical absorption spectroscopy (DOAS) observations.
Absorption by a dominant well-mixed gas like O$_2$ actually contains information on the vertical structure of the scattering particles, a clearly desirable piece of information in cloud and aerosol studies.
The intermediate property that links the O$_2$ DOAS to the scattering layers is the distribution (e.g., mean and variance) of the path length cumulated by the reflected sunlight through multiple scatterings.
A multi-angle measurement of water vapor absorption can be used to reconstruct approximately the water-vapor column.
Water vapor, aerosols (more precisely, cloud condensation nuclei) and convective motion are what produce clouds in the first place; so it is highly desirable to assay the water vapor column at the same time and place as the convectively-driven clouds.

In closing, all of the above instrumentation has to be vetted for Size, Weight and Power (SWAP) available on a NanoSat.
Also, a rigorous tradeoff study is in order for each one to obtain the required signal-to-noise ratio (SNR) to achieve the science and application mission goals, by co-adding pixels and adjusting exposure time, if necessary.

\section{Engineering and Feasibility}


\begin{table*}
\renewcommand{\arraystretch}{1.3}
\caption{\bf Summary of the NanoSat Components}
\label{table:spacecraft}
\centering
\begin{tabular}{|c|c|c|c|c|c|c|}
\hline
\bfseries Component & \bfseries COTS Model & \bfseries Size [mm] & \bfseries Mass [g] & \bfseries Power [W] & \bfseries Performance & \bfseries TRL\\
\hline\hline
Propulsion & VACCO MIPS & 1U & 1245 & 0.25 (standby)& 250 Ns (total impulse) & 9 \\
& & & & 10 (max) & & \\
               & Aerojet MPS-120 & 1U & 1600 & N/A & 800 Ns (total impulse)  & 9 \\
               \hline
ADCS & BCT XACT-100 & 0.5U & 1520 & N/A & 0.003 deg (pointing accuracy) & 9\\
& & & & & 100 mNms (momentum capacity) & \\
\hline
GPS Receiver & Novatel OEM7600 & 35x55x13 & 31 & 1.3 & 1 cm (position accuracy, RTK) & 9 \\
\hline
Communication & NanoAvionics & 56x33x6.5 & 7.5 & 3 & 38.4 kbps (data-rate) & 9\\
 & SatCOM UHF &  &  &  & & \\
 & TESAT CubeLCT & 0.3U & 360 & 8 & 100 Mbps (LEO to ground) & 6\\
\hline
Computation     & Qualcomm & 75x26 & N/A & N/A & Quad-core 2.15 GHz Kryo (CPU) & 7\\
 & Snapdragon Flight & & & & 4GB LPDDR4 RAM  & \\
 & Nvidia Jetson TX2 & 87x50 & 85 & 7.5 (nominal) & 256 Nvidia CUDA cores (GPU) & 4\\
 & & & & 15 (max) & Dual-core Nvidia Denver (CPU)  & \\
  & & & & & 8GB LPDDR4 RAM  & \\
\hline
Power & EnduroSat Solar Array & 3U & 270 & 8.4 & 29.5\% (power efficiency) & 9 \\
 & (Deployable) &  &  &  & 8.4 W (max power per side) & \\
\hline
\end{tabular}
\end{table*}

The GEOSCAN constellation consists of a swarm of NanoSats that are identical in geometrical shape and configuration. In this section, we first describe the common hardware components of the NanoSat spacecraft that form basis of the constellation. Then we focus on how the GEOSCAN will coordinate the swarm of NanoSats to enable adaptive sensing of a natural phenomena and identify software gaps that need to be bridged to enable the capability.

\subsection{Spacecraft}
The NanoSat will have form factor of a standard 6U CubeSat platform that can be readily deployed from standard dispensers such as Poly-Picosatellite Orbital Deployer (P-POD) \cite{chin2008cubesat}. The NanoSat will utilize commercial-off-the-self (COTS) components for reduced cost and feasibility for production of hundreds required for the constellation. Table \ref{table:spacecraft} lists main system components that will comprise 6U NanoSat spacecraft. The selection is based on size, weight, and power of each components in relation to the performance each can provide to meet the science requirements.

The propulsion unit will use cold/warm propulsion units as a baseline. The cold/warm propulsion are advantageous as they are simple in mechanism, involves no chemical reaction, non-toxic, low cost, and already been flight-proven and highly matured for NanoSats \cite{nasaSoA2021}. C-POD and MIPS from VACCO are two possible COTS candidates for propulsion unit with the former providing total impulse of 186Ns and the latter 250Ns with 1U unit. These, however, provides limited propulsion that will restrict the number of active formation reconfiguration maneuvers that can be performed. To overcome the limited propulsion of cold/warm units, a hydrazine propulsion system can be considered as an alternative. Hydrazine propulsion is corrosive, toxic, and its vapor pressure requires overhead that may not be ideal for NanoSats. However, it can provide several fold greater total impulse (i.e.,  MPS-120 from Aerojet has total impulse of over 0.8~kNs/2~kNs with 1U/2U unit, respectively) that will enable more frequent formation reconfiguration and/or longer duration of mission operation.

The Attitude Determination and Control System (ADCS) will use XACT integrated unit from Blue Canyon Technologies. XACT integrated ADCS unit is TRL 9 with flight-proven in its usage in both MarCo and ASTERIA. It houses 3 reaction wheels and 3 magnetorquers for actuation, and 1 star tracker and 3-axis magnetometer for sensing. The actuators and sensors together provide pointing accuracy 0.003~deg for two axes and 0.007~deg for the 3$^\text{rd}$ axis. The level of pointing accuracy can sufficiently accommodate up to 50~m pixel spatial resolution from on-board cameras. 

The NanoSat is situated in low-Earth orbit and will hence utilize GNSS signals to perform orbit determination and localization. The on-board GPS receivers are matured for small spacecraft and several COTS GPS receivers chips are available. Here, we will use OEM7600 from NovaTel that is able to provide centimeter-level real-time positioning when combined with RTK.

The communication module has matured for CubeSat in the VHF and UHF frequency bands with several COTS products that are TRL 9. The NanoSat will use SatCOM UHF transiever from NanoAvionics that operates in frequency range of 395-440~MHz and attain max data rate of 38.4~kbps. It uses half duplex RF architecture and is comparably low in weight and power consumption with respect to other COTS products. However, the data-rate provided by UHF transceivers may not meet the needs for higher bandwidth to transmit high-resolution measurement data down to the ground. As an alternative, laser communication can become a viable option that is quickly maturing to TRL 7 and above. Among several under development products, CubeLCT from TESAT is one of the smallest laser communication transmitter that can provide up to 100~Mbps downlink from LEO in a compact 0.3U form factor which the NanoSat can utilize.

The NanoSat will require sufficiently high-computing unit to enable real-time computation of planning algorithms that is crucial to autonomy capabilities of GEOSCAN constellation. On-board computers that have flight heritage falls short of the required computing power. Two COTS units are viable options for their computing power, despite having little or no flight heritage. Snapdragon flight platform from Qualcomm has been successfully powered autonomous flight of the Ingenuity and can be utilized to power the computing needs of the NanoSat. Nvidia's Jetson TX2 packs more computing power and better suited to run AI/ML algorithms being equipped with GPU. It can be a favorable option if it can be engineered to be space-certified in the near future.

Power is also an important component that includes unit for power generation, storage, and management. Among different means of power generation, solar power is most widely used and has high maturity for small spacecraft. The NanoSat will utilize deployable solar panel from EnduroSat that is space-qualified and has flight heritage. The 3U deployable solar array can generate up to max power of 8.4~W for each side. The module provides both series or parallel connection and a minimum of two solar arrays will be required to provide the max power 16.8W to cover the power requirements of other on-board components. Having more than two solar array may also be considered if different modules need to operate concurrently throughout the mission.


\subsection{Formation Initialization and Reconfiguration}

\begin{figure}[t!]
	\centering
	\includegraphics[width=1.0\linewidth]{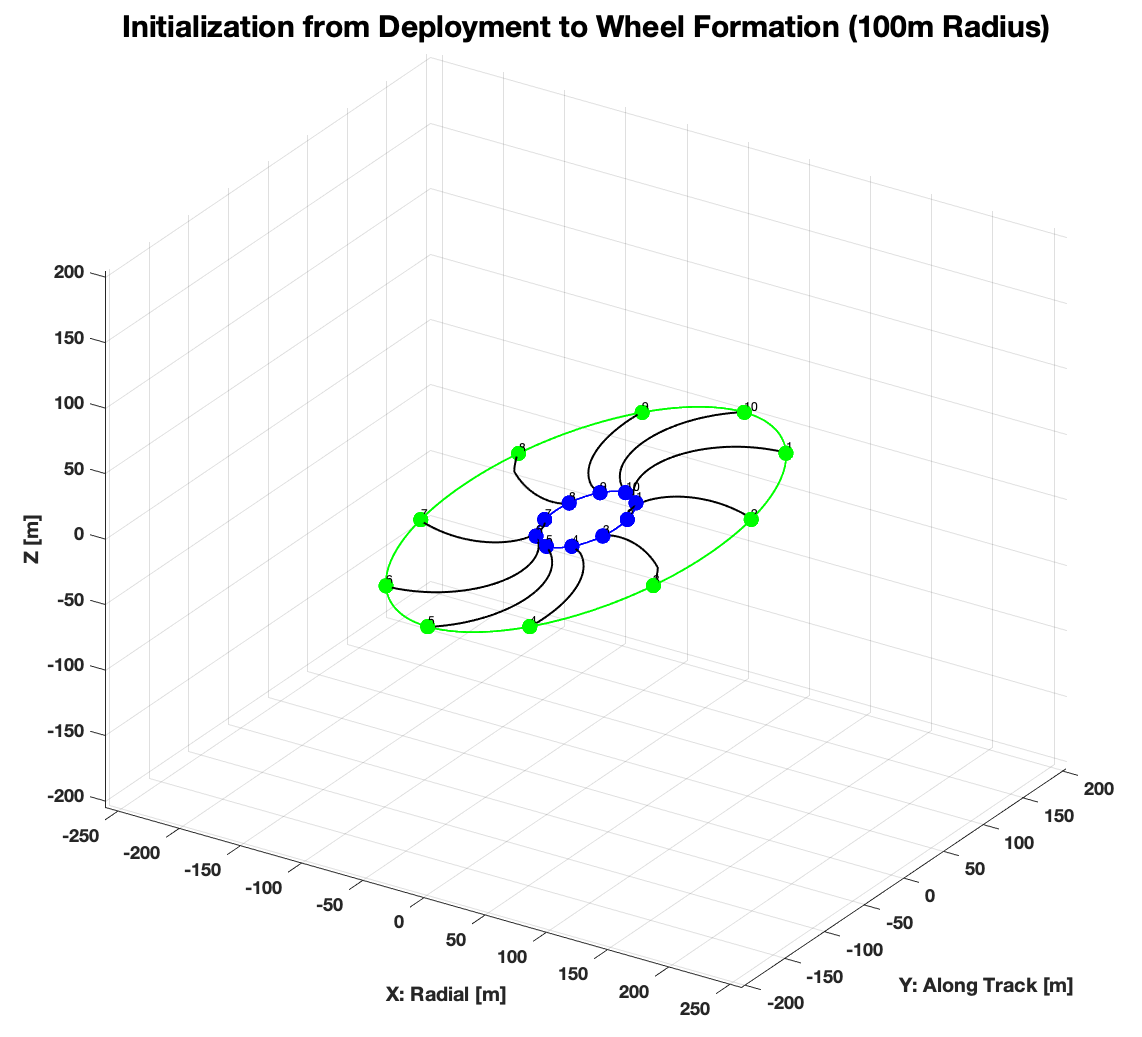}
	\caption{Each fleet in the GEOSCAN constellation will go through initialization to a wheel formation after deployment. On-board optimization-based algorithm will autonomously compute minimum fuel trajectories for this maneuver without ground-in-the-loop intervention.}
	\label{fig:formation_init}
\end{figure}

A fleet of NanoSat will be deployed from dispensers such as Poly-Picosatellite Orbital Deployer (P-POD) or Nanoracks CubeSat Deployer (NRCSD). Once deployed, the fleet will need to initialize itself to a wheel formation of a desirable radius for scientific measurements.  Figure \ref{fig:formation_init} shows an example of the fleet's initialization maneuver to form itself to a wheel of radius 100~m. On-board planning algorithm that utilizes recent advances in optimization \cite{morgan2016swarm}, will compute optimal trajectories for this transfer autonomously without ground-in-the-loop intervention. The planning algorithm will guarantee that the generated trajectories are minimal in fuel cost and safe from collision. Note that the example trajectories in Figure \ref{fig:formation_init} are shown for fleet of 10 NanoSat in the Local-Vertical Local-Horizontal (LVLH) frame with origin at the center of the wheel. The LVLH frame follows convention from Hill's frame with $x$-axis pointing radially away from Earth, $y$-axis in the direction of orbital velocity, and $z$-axis in the direction of angular momentum.

\begin{figure}[h!]
	\centering
	\includegraphics[width=1.0\linewidth]{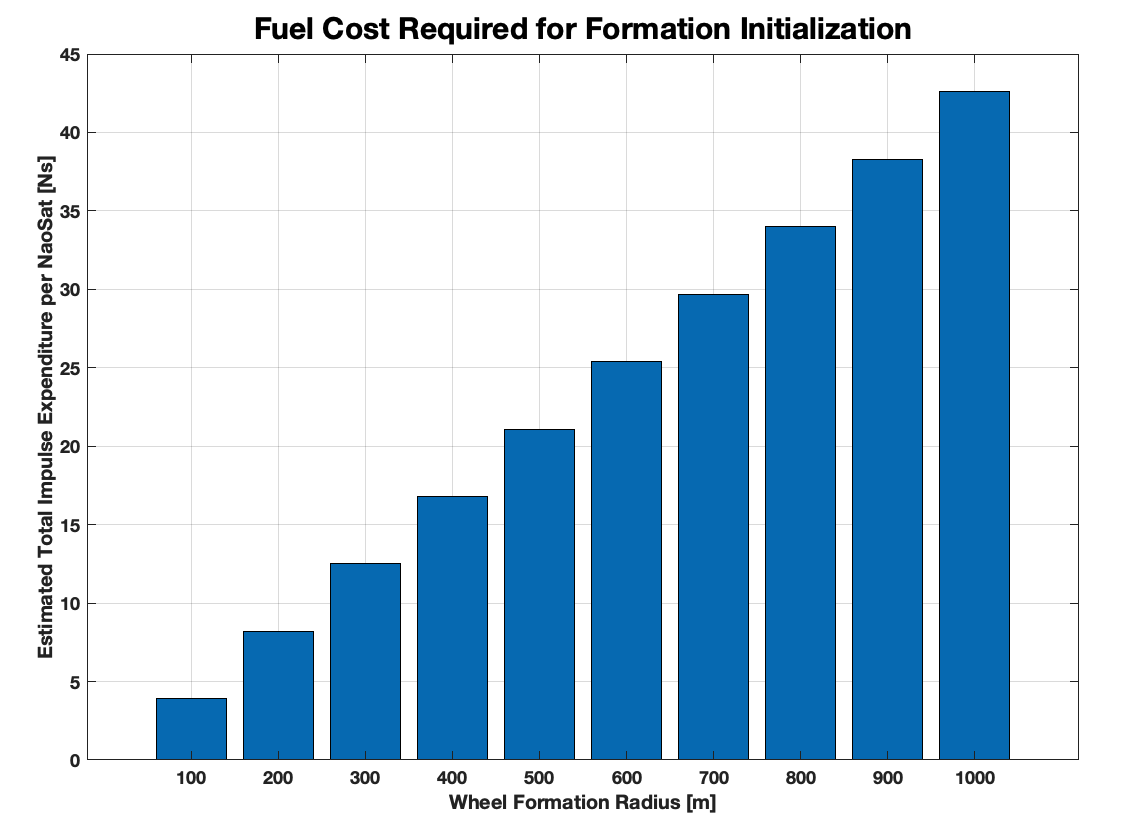}
	\caption{The simulated fuel expenditure of initial formation is shown for varying wheel radius. The transfer time limit of 15~min (roughly 1/6 of orbital period) was imposed throughout the simulation with spacecraft mass of 10~kg assumed. The estimated fuel cost linearly increases as the desired wheel radius enlarges.}
	\label{fig:formation_init_cost}
\end{figure}

Despite the effort of optimization based algorithm to compute minimum fuel trajectories, the initialization maneuver to form wheel has non-negligible cost associated with it. The two parameters that affect the fuel cost are the wheel formation radius and time limit to complete the transfer. Generally speaking, fuel cost increases when the desired wheel radius is small and the time to transfer is short. Figure \ref{fig:formation_init_cost} shows estimated fuel cost for forming wheel of radius ranging from 100~m to 1000~m, computed from simulation with transfer time limit of 15~min. The simulation result shows linear increase in the initialization fuel cost with respect to the desired wheel radius ranging from 3.9~Ns for 100~m to 42.6~Ns for 1000~m.

\begin{figure}[t!]
	\centering
	\includegraphics[width=1.0\linewidth]{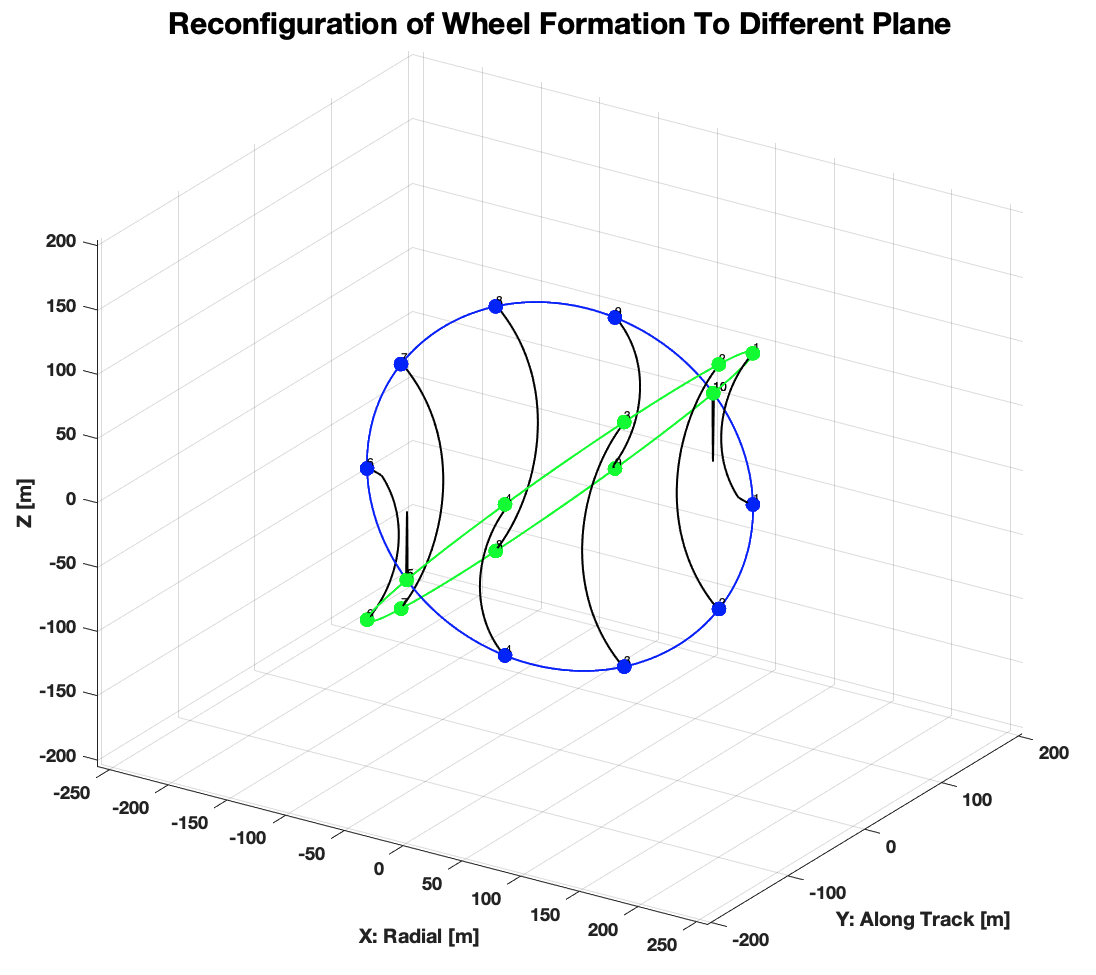}
	\caption{The simulated fuel expenditure of initial formation is shown for varying wheel radius. The transfer time limit of 15~min (roughly 1/6 of orbital period) was imposed throughout the simulation with spacecraft mass of 10~kg assumed. The estimated fuel cost linearly increases as the desired wheel radius enlarges.}
	\label{fig:formation_reconfig}
\end{figure}

Once a fleet is initialized to a wheel formation, it will maintain its formation till a reconfiguration becomes necessary to maximize science return of the observation. Similarly to initialization, the maneuver to reconfigure the formation will also be computed autonomously on-board using planning algorithms without ground-in-the-loop guidance. Figure~\ref{fig:formation_reconfig} shows an example of a fleet changing the wheel formation onto a different plane that is perpendicular to the current plane, while keeping the radius constant at 100~m. The fuel cost of reconfiguration will vary depending on the desired formation shape to transform into. The simulated experiments show that reconfiguration will, on average, require around 2-fold fuel cost to that of initialization for a given wheel radius.

Assuming the fleet operate in a wheel formation of radius 100~m, this corresponds to $\sim$8~Ns expenditure in the limited total impulse budget that each NanoSat has. Assuming cold gas propulsion unit is selected (e.g., VACCO MIPS) and no more than a third of the total impulse is allocated for reconfiguration, the fleet can perform less than 10 reconfiguration throughout the mission lifetime. Give the limited number of reconfiguration that can be performed, a fleet must carefully choose when to reconfigure by weighing the expected increase in the science return to the the reduction of the fuel budget.

\subsection{Formation Upkeep}

The relative positions of the NanoSat will drift over time and correction must be made to maintain a desirable wheel formation. The along-track drift rate of a NanoSat relative to the center of the wheel is expressed as \cite{alfriend2009spacecraft}:
\begin{equation}
\dot{y}_{drift} = -\frac{3n}{2a}(2\rho^2_x + 2\rho^2_y + \rho^2_z + 6\rho_x\rho_y\cos\alpha_x + 3\rho^2_x \cos 2\alpha_x)
\end{equation}

The $\rho_x, \rho_y, \rho_z, \alpha_x$ are parameters that describe relative orbit of a spacecraft in magnitude-phase form, $a$ is altitude, and $n$ is mean motion of the orbit. For a projected circular orbit (PCO), a linear estimate of along-track drift per orbit can be derived \cite{alfriend2009spacecraft} and given by:

\begin{equation}
y_{drift} = -\frac{9\pi \rho^2}{2a}(2+\cos 2\alpha_x)
\end{equation}

The drift is proportional to the square of the wheel radius and hence significantly larger correction must be made if the desirable radius of the formation increases. The GEOSCAN constellation will operate at an altitude that is slightly lower than ISS (400~km). Assuming a fleet operates in the wheel formation of radius 100~m, this corresponds to $\sim$20~cm of drift per one orbit. In comparison, formation keeping of 1000~m radius will require $\sim$20~m drift correction every orbit that will expend significant amount of the fuel.

Based on the simulation, the fuel cost to maintain wheel formation of 100~m radius is computed to be $\sim$31~Nm for the duration of one year. This will allow around 5 years of operation even when the NanoSat is equipped with cold gas propulsion unit, and this duration can be extended to more than a decade if an alternative propulsion unit with higher total impulse (e.g., hydrazine propulsion) is utilized. However, wheel formation of 1000~m radius is two orders of magnitude more expensive to maintain and can quickly deplete the fuel budget within few weeks.

\subsection{Coordinated Adaptive Pointing}

The previous discussed formation initialization, reconfiguration, and maintenance required positional adjustments that needs to be planned out. Once fleets are in the desired formation and positions are set, planning out attitudes of NanoSats to adjust sensor pointing directions become important during the observation phase. The NanoSats within a fleet and fleets within a string-of-pearls of the GEOSCAN constellation will have to be coordinate closely to enable multi-angular, temporal, and spectral observation of an evolving natural phenomena. 

One of the key software challenge will be developing computationally efficient methodology for generating optimal attitude plan that coordinates pointing distributed sensors to maximize the scientific value of the measurements. This will involve combination of high-level scheduling and low-level attitude planning/controlling. The high-level scheduling will make continuous selection/de-selection of NanoSats to participate in the coordinated observation given the state of each NanoSats. The scheduling problem for constellation is often formulated as optimization problem, especially in the form of mixed integer linear program (MILP) in the literature \cite{nag2016simulation}, \cite{nag2018scheduling} \cite{nag2019autonomous}, \cite{sanchez2018starling1}. However, the shear size of the constellation and the non-convexity nature of the problem makes the state-of-the-art methodology computationally intractable for real-time applications.

This is a critical technology gap that needs to be bridged to enable real-time coordinated measurement of rapidly evolving phenomena using distributed sensors in the constellation. A recent research direction that holds promise is methods that leverage both the machine learning and optimization to tackle the computational complexity. Deep neural net, when adequately trained using appropriate dataset, is shown to have potential to very quickly generate near optimal solutions to non-convex problems \cite{yun2020multi}. Similar approach can be taken to tackle constellation scheduling problem for coordinated observation.

Once high-level scheduling plan is generated, the low-level attitude planning and controlling can be computed on-board on each of the NanoSats. The state-of-art methodologies are sufficiently fast to operate both attitude planning and controlling in a high frequency to provide accurate sensor pointing with minimum torque actuation. An example of multi-spacecraft attitude planning and controlling that can be utilized is detailed in \cite{nakka2021information}.


\section{Discussion}

The previous section focused on the different engineering components of the active steering capabilities that are essential to GEOSCAN mission. Nevertheless, there are several other aspects of the mission concept that are important to consider. We list here certain elements of the mission that will require in-depth analysis to concretize the concept further.

One of the key element will be choosing how many of string-of-pearls formations to have and which orbits each will be assigned to. The Figure \ref{fig:mission_concept} shows GEOSCAN with one string-of-pearls formation (red) for illustration, but multiple string-of-pearls is possible and will likely be desirable to diversify the vantage points of the scientific measurements. Several factors will weigh in this consideration, including Earth coverage, sun angle, communication mesh, ease of orbit insertion, cost, redundancy, and many more. 

Another key question to be answered will be how to optimize the replenishment strategy to enable continuous operation of the GEOSCAN in the presence NanoSats failures. GEOSCAN will inherent be highly robust to failures with the constellation having swarm of NanoSats that can distribute its operational functionalities. However, the limited lifetime of NanoSats pose a significant challenge to its continuous operation going beyond several years. To this end, the failure rate of the NanoSats and its impact on the performance of GEOSCAN must be quantified and an appropriate replenishment strategy must be developed to ensure continuous operation without performance degradation. A strategy to safely de-orbit NanoSats nearing end of the lifetime must also be designed.

The NanoSats within GEOSCAN will be heterogeneous, not only in the scientific instrument each carry, but in the primary functionality it provides. Due to the limited form factor, it will likely be infeasible to perform sensing, actuation, communication, and computation concurrently in a single NanoSat. Each fleet will therefore have members that do not participate in sensing, but instead, dedicated to communication and computation. The chief NanoSat at the center of the wheel is an ideal candidate for this but it is possible to have more than one NanoSat within a fleet for this purpose. Selection of ratio between sensing and non-sensing units within a fleet will need to be chosen based on the needs.

The NanoSats that are used for sensing, will undergo continuous attitude adjustments while it gathers scientific data of a target phenomena. The ADCS will primary use reaction wheels to perform the required orientation changes, but the momentum capacity is often limited for miniaturized ADCS for NanoSats. As a result, each NanoSat will required to perform de-saturation maneuver on a regular interval, to shed off the momentum from saturated reaction wheels. During de-saturation maneuver, the NanoSat will not only consume fuel but will have no control authority over its pointing and become unavailable for sensing. An online scheduling algorithm that plans out sensing and de-saturating phases of NanoSats must be looked into to prevent undesirable interruption in the measurement data of the target.

Data volume management will also be an important aspect to consider. The raw data generated by each sensor on-board different NanoSats will quickly outgrow the storage and communication bandwidth of the system. How to downlink the gathered data to the ground for CCT reconstruction will be a hard challenge to be resolved. The solution will have to come from both the hardware and software enhancements. A push towards optical communication and AI/ML based sampling/compression algorithms will likely hold key to this.

\section{Conclusion}

In this paper, we presented a future Earth observation mission concept named GEOSCAN that utilizes constellation of NanoSats to provide multi-dimensional scientific measurements of a natural atmospheric phenomena from low Earth orbit. The key science-enabling driver for the GEOSCAN mission is the emerging field of 3D computed cloud tomography (CCT) that uses multi-angular and spectrally-resolved measurements to reconstruct 3D structure of the clouds. The key engineering capability of the GEOSCAN constellation is autonomous scheduling and steering of the NanoSats for coordinated and distributed observation from multi-vantage points. The key benefit is in having more accurate characterization of the clouds, which can lead to significant reduction of uncertainties in the global climate models (GCMs) to improve our understanding of, and response to the climate crisis. Our work discussed various elements of the proposed mission concept from both scientific and engineering perspectives. 

Science side, we outlined the types of remote Earth observation measurements that GEOSCAN enables beyond the state-of-the-art, and how such measurements translate to improvements in CCT that can lead to reduction in uncertainty of GCMs. We also described some applied science potentially enabled by GEOSCAN, namely, a CCT-based characterization of cloud-like atmospheric phenomena that cause natural hazards (specifically, wildfire smoke and volcanic ash plumes in their original opaque phase). On the engineering side, we investigated feasibility of the concept starting from hardware components of the NanoSat that form the basis of the constellation, and focused on the active steering capability of the GEOSCAN with algorithmic approaches that enable coordination. We identified technology gaps that need to be bridged, and discussed other aspects of the mission that require in-depth analysis to further mature the mission concept.

\acknowledgments
This research was carried out at the Jet Propulsion Laboratory, California Institute of Technology, under a contract with the National Aeronautics and Space Administration (80NM0018D0004). We thank Linda Forster, Aviad Levis, Yoav Schechner for valuable scientific discussions and Ramtin Madani, Kyongsik Yun, Alex Sabol, Muhammad Adil for AI/ML technical developments that the proposed concept leverages. 

\bibliographystyle{IEEEtran}

\begin{thebibliography}{10}

\bibitem{ipcc2021climate}
V.~Masson-Delmotte, P.~Zhai, A.~Pirani, S.L.Connors, C.~P\'{e}an, S.~Berger,
  N.~Caud, Y.~Chen, L.~Goldfarb, M.~Gomis, M.~Huang, K.~Leitzell, E.~Lonnoy,
  J.~Matthews, T.~Maycock, T.~Waterfield, O.~Yelek\c{c}i, R.~Yu, and B.~Zhou,
  ``{Climate Change 2021: The Physical Science Basis. Contribution of Working
  Group I to the Sixth Assessment Report of the Intergovernmental Panel on
  Climate Change},'' \emph{Intergovernmental Panel on Climate Change (IPCC),
  Cambridge University Press, Cambridge}, 2021.

\bibitem{stephens2018cloudsat}
G.~Stephens, D.~Winker, J.~Pelon, C.~Trepte, D.~Vane, C.~Yuhas, T.~L\'{e}cuyer,
  and M.~Lebsock, ``{CloudSat and CALIPSO within the A-Train: Ten years of
  actively observing the Earth system},'' \emph{Bulletin of the American
  Meteorological Society}, vol.~99, no.~3, pp. 569--581, 2018.

\bibitem{nakajima1990determination}
T.~Nakajima and M.~D. King, ``Determination of the optical thickness and
  effective particle radius of clouds from reflected solar radiation
  measurements. {Part I: T}heory,'' \emph{Journal of Atmospheric Sciences},
  vol.~47, no.~15, pp. 1878--1893, 1990.

\bibitem{NASEM2018}
{National Academies of Sciences, Engineering, and Medicine}, \emph{Thriving on
  Our Changing Planet, A Decadal Strategy for Earth Observation from
  Space}.\hskip 1em plus 0.5em minus 0.4em\relax Washington (DC): The National
  Academies Press, 2018.

\bibitem{Levis_etal2015}
A.~Levis, Y.~Y. Schechner, A.~Aides, and A.~B. Davis, ``Airborne
  three-dimensional cloud tomography,'' in \emph{Proceedings of the IEEE
  International Conference on Computer Vision}, 2015, pp. 3379--3387.

\bibitem{Levis_etal2017}
A.~Levis, Y.~Y. Schechner, and A.~B. Davis, ``Multiple-scattering microphysics
  tomography,'' in \emph{Proceedings of the IEEE Conference on Computer Vision
  and Pattern Recognition}, 2017, pp. 6740--6749.

\bibitem{Levis_etal2020}
A.~Levis, Y.~Y. Schechner, A.~B. Davis, and J.~Loveridge, ``Multi-view
  polarimetric scattering cloud tomography and retrieval of droplet size,''
  \emph{Remote Sensing}, vol.~12, no.~17, p. 2831, 2020.

\bibitem{martin2014adjoint}
W.~Martin, B.~Cairns, and G.~Bal, ``Adjoint methods for adjusting
  three-dimensional atmosphere and surface properties to fit
  multi-angle/multi-pixel polarimetric measurements,'' \emph{Journal of
  Quantitative Spectroscopy and Radiative Transfer}, vol. 144, pp. 68--85,
  2014.

\bibitem{martin2018demonstration}
W.~G. Martin and O.~P. Hasekamp, ``A demonstration of adjoint methods for
  multi-dimensional remote sensing of the atmosphere and surface,''
  \emph{Journal of Quantitative Spectroscopy and Radiative Transfer}, vol. 204,
  pp. 215--231, 2018.

\bibitem{doicu2021cloud}
A.~Doicu, A.~Doicu, D.~Efremenko, and T.~Trautmann, ``Cloud tomographic
  retrieval algorithms. {I: S}urrogate minimization method,'' \emph{Journal of
  Quantitative Spectroscopy and Radiative Transfer}, p. 107954, 2021.

\bibitem{Levis_etal2021}
A.~Levis, A.~B. Davis, J.~R. Loveridge, and Y.~Y. Schechner, ``{3D} cloud
  tomography and droplet size retrieval from multi-angle polarimetric imaging
  of scattered sunlight from above,'' in \emph{Polarization Science and Remote
  Sensing X}, vol. 11833.\hskip 1em plus 0.5em minus 0.4em\relax International
  Society for Optics and Photonics, 2021, pp. 11\,833--05.

\bibitem{diner2013airborne}
D.~J. Diner, F.~Xu, M.~J. Garay, J.~V. Martonchik, B.~E. Rheingans, S.~Geier,
  A.~Davis, B.~Hancock, V.~Jovanovic, M.~Bull \emph{et~al.}, ``{The Airborne
  Multiangle SpectroPolarimetric Imager (AirMSPI): A} new tool for aerosol and
  cloud remote sensing,'' \emph{Atmospheric Measurement Techniques}, vol.~6,
  no.~8, pp. 2007--2025, 2013.

\bibitem{evans1998spherical}
K.~F. Evans, ``The spherical harmonics discrete ordinate method for
  three-dimensional atmospheric radiative transfer,'' \emph{Journal of the
  Atmospheric Sciences}, vol.~55, no.~3, pp. 429--446, 1998.

\bibitem{schilling2019cloudct}
K.~Schilling, Y.~Y. Schechner, and I.~Koren, ``{CloudCT}-computed tomography of
  clouds by a small satellite formation,'' in \emph{Proceedings of the 12th IAA
  symposium on Small Satellites for Earth Observation}, vol.~6, 2019, p.~7.

\bibitem{tzabari2021cloudct}
M.~Tzabari, V.~Holodovsky, O.~Shubi, E.~Eitan, O.~Altaratz, I.~Koren,
  A.~Aumann, K.~Schilling, and Y.~Y. Schechner, ``{CloudCT 3D volumetric
  tomography: C}onsiderations for imager preference, comparing visible light,
  short-wave infrared, and polarized imagers,'' in \emph{Polarization Science
  and Remote Sensing X}, vol. 11833.\hskip 1em plus 0.5em minus 0.4em\relax
  International Society for Optics and Photonics, 2021, p. 1183304.

\bibitem{bandyopadhyay2016review}
S.~Bandyopadhyay, R.~Foust, G.~P. Subramanian, S.-J. Chung, and F.~Y. Hadaegh,
  ``Review of formation flying and constellation missions using
  nanosatellites,'' \emph{Journal of Spacecraft and Rockets}, vol.~53, no.~3,
  pp. 567--578, 2016.

\bibitem{forster2021toward}
L.~Forster, A.~B. Davis, D.~J. Diner, and B.~Mayer, ``Toward cloud tomography
  from space using {MISR and MODIS: L}ocating the ``veiled core'' in opaque
  convective clouds,'' \emph{Journal of the Atmospheric Sciences}, vol.~78,
  no.~1, pp. 155--166, 2021.

\bibitem{chin2008cubesat}
A.~Chin, R.~Coelho, R.~Nugent, R.~Munakata, and J.~Puig-Suari, ``{CubeSat: T}he
  pico-satellite standard for research and education,'' in \emph{AIAA Space
  2008 Conference \& Exposition}, 2008, p. 7734.

\bibitem{nasaSoA2021}
``{State-of-the-Art Small Spacecraft Technology},'' NASA Ames Research Center,
  report, 2021.

\bibitem{morgan2016swarm}
D.~Morgan, G.~P. Subramanian, S.-J. Chung, and F.~Y. Hadaegh, ``Swarm
  assignment and trajectory optimization using variable-swarm, distributed
  auction assignment and sequential convex programming,'' \emph{The
  International Journal of Robotics Research}, vol.~35, no.~10, pp. 1261--1285,
  2016.

\bibitem{alfriend2009spacecraft}
K.~T. Alfriend, S.~R. Vadali, P.~Gurfil, J.~P. How, and L.~Breger,
  \emph{Spacecraft formation flying: {D}ynamics, control and navigation}.\hskip
  1em plus 0.5em minus 0.4em\relax Elsevier, 2009, vol.~2.

\bibitem{nag2016simulation}
S.~Nag, C.~K. Gatebe, and T.~Hilker, ``Simulation of multiangular remote
  sensing products using small satellite formations,'' \emph{IEEE Journal of
  Selected Topics in Applied Earth Observations and Remote Sensing}, vol.~10,
  no.~2, pp. 638--653, 2016.

\bibitem{nag2018scheduling}
S.~Nag, A.~S. Li, and J.~H. Merrick, ``Scheduling algorithms for rapid imaging
  using agile {C}ubesat constellations,'' \emph{Advances in Space Research},
  vol.~61, no.~3, pp. 891--913, 2018.

\bibitem{nag2019autonomous}
S.~Nag, A.~S. Li, V.~Ravindra, M.~S. Net, K.-M. Cheung, R.~Lammers, and
  B.~Bledsoe, ``Autonomous scheduling of agile spacecraft constellations with
  delay tolerant networking for reactive imaging,'' in \emph{International
  Conference on Automated Planning and Scheduling}, 2019.

\bibitem{sanchez2018starling1}
H.~Sanchez, D.~McIntosh, H.~Cannon, C.~Pires, J.~Sullivan, B.~O'Connor, and
  S.~D'Amico, ``Starling1: {Swarm Technology Demonstration},'' in \emph{32nd
  Annual Small Satellite Conference}, 2018.

\bibitem{yun2020multi}
K.~Yun, C.~Choi, R.~Alimo, A.~B. Davis, L.~Forster, A.~Rahmani, M.~Adil, and
  R.~Madani, ``{Multi-Agent Motion Planning using Deep Learning for Space
  Applications},'' in \emph{ASCEND 2020}, 2020, p. 4233.

\bibitem{nakka2021information}
Y.~K.~K. Nakka, W.~H{\"o}nig, C.~Choi, A.~Harvard, A.~Rahmani, and S.-J. Chung,
  ``Information-based guidance and control architecture for multi-spacecraft
  on-orbit inspection,'' in \emph{AIAA Scitech 2021 Forum}, 2021, p. 1103.

\end{thebibliography}

\thebiography

\begin{biographywithpic}
{Changrak Choi}{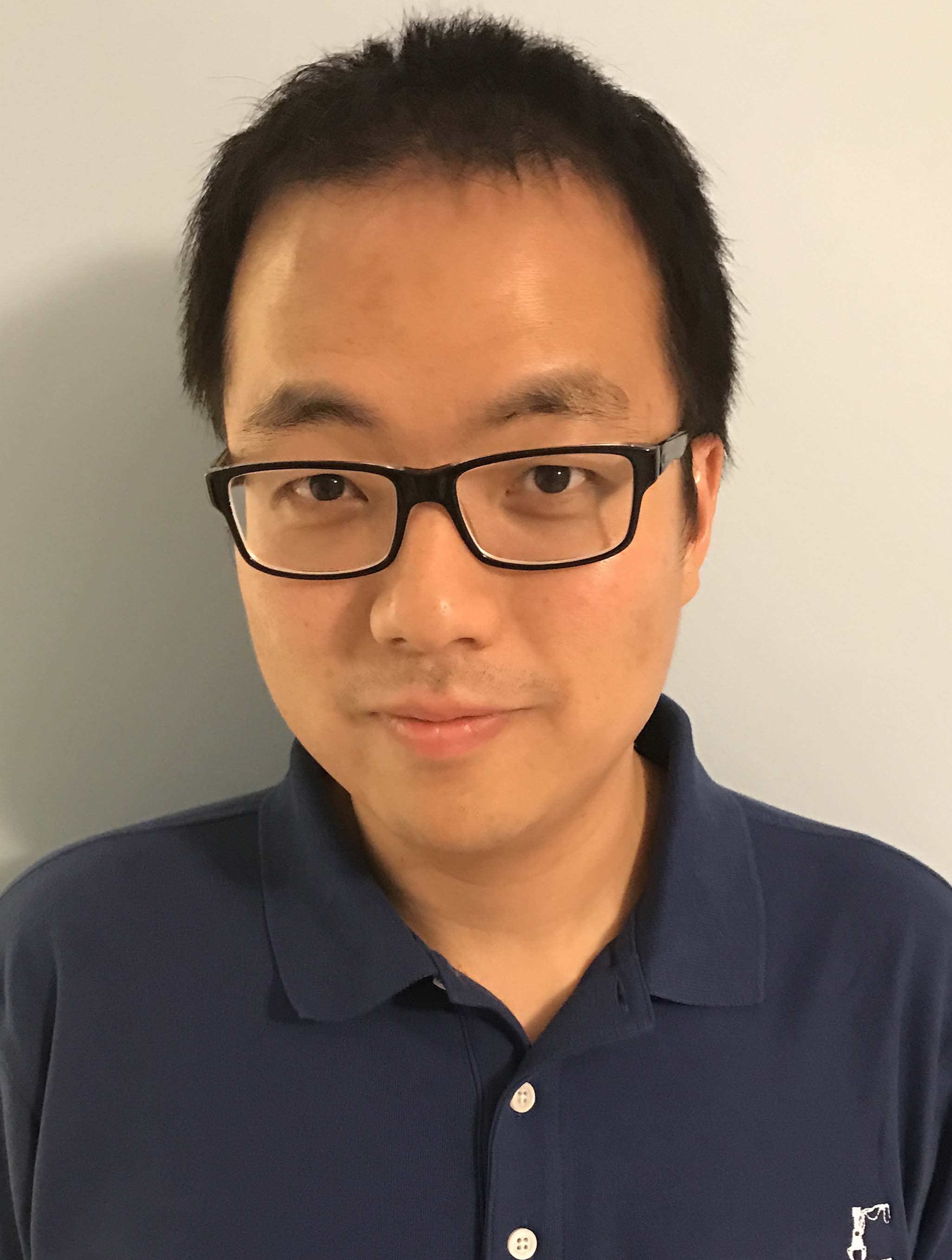}
is a Robotics Technologist in the Multi-Agent Autonomy group at JPL. His research focuses on trajectory and motion planning for autonomous systems that are dynamically interesting, drawing upon algorithms, optimization, and controls. The systems of interest range from ground to on-water and aerial vehicles as well as soft robots, with a special interest in multi-spacecraft systems. Changrak received his PhD from MIT as a member of Laboratory for Information and Decision Systems (LIDS) in Aerospace Robotics and Embedded Systems group. He earned MS and BS, both in Mechanical Engineering, from MIT and Seoul National University respectively.
\end{biographywithpic} 

\begin{biographywithpic}
{Anthony Davis}{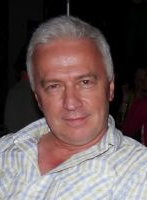}
holds a BSc in Physics/Astrophysics from Universit\'{e} Pierre et Marie Curie (1977), an MS in Physics/Astronomy from Universit\'{e} de Montr\'{e}al (1981), and a PhD in Physics from McGill University (1992). He has worked for over three decades on innovative methods in cloud remote sensing from ground-, aircraft- and space-based platforms, as well as on the related problem of how clouds impact the Earth's energy and hydrological cycles. He has held positions successively at NASA Goddard Space Flight Center in Greenbelt, Maryland, DOE's Los Alamos National Laboratory in New Mexico, and now at NASA’s Jet Propulsion Laboratory in Pasadena, Ca. Dr. Davis specializes in the challenges posed by the 3D shapes and internal structures of real clouds to which he brings to bear his expertise in theoretical and computational radiative transfer in application to physics-based modeling and exploitation of the spatial, directional, spectral and/or polarimetric signals from clouds. Anthony's interests have recently grown to include remote sensing of aerosol/surface properties, particularly in the vicinity of clouds, or other spatially complex situations using multi-pixel/multi-angle spectral data, potentially with polarization diversity. Finally, he supports JPL planetary mission formulation projects in matters of imaging through gases, hazes and clouds.
\end{biographywithpic}

\end{document}